\documentclass[11pt]{article}
\usepackage{moriond,epsfig}
\usepackage{subfigure}

\def\Journal#1#2#3#4{{#1} {\bf #2}, #3 (#4)}

\def\PRL{\em Phys. Rev. Lett.}
\def\PRD{{\em Phys. Rev.} D}

\def\HEP{\em J. High Energy Phys.}
\def\CPC{\em Comput. Phys. Commun.}

\def\be{\begin{equation}}
\def\ee{\end{equation}}
\def\bea{\begin{eqnarray}}
\def\eea{\end{eqnarray}}

\begin{document}
\vspace*{4cm}
\title{MEASUREMENT OF THE W-BOSON HELICITY FRACTIONS IN TOP-QUARK DECAYS AT CDF}

\author{ T. CHWALEK }

\address{Institut f\"ur Experimentelle Kernphysik, University of Karlsruhe,\\
Wolfgang-Gaede-Str.1, 76131 Karlsruhe, Germany}

\maketitle\abstracts{
We present a measurement of the fractions $F_{0}$ and $F_{+}$ of longitudinally polarized and right-handed $W$ bosons in top-quark decays using data collected with the CDF~II detector. The data set used in the analysis corresponds to an integrated luminosity of approximately 955~$\rm{pb}^{-1}$. We select $t\bar{t}$ candidate events with one lepton, at least four jets, and missing transverse energy. Our helicity measurement uses the decay angle $\theta^{*}$, which is defined as the angle between the momentum of the charged lepton in the $W$ boson rest-frame and the $W$ momentum in the top-quark rest-frame. The $\cos\theta^{*}$ distribution in the data is determined by full kinematic reconstruction of the $t\bar{t}$ candidates. We find $F_{0}=0.59\pm 0.12(\rm stat)^{+0.07}_{-0.06}(\rm syst)$ and $F_{+}=-0.03\pm 0.06(\rm stat)^{+0.04}_{-0.03}(\rm syst)$, which is consistent with the standard model prediction. We set an upper limit on the fraction of right-handed $W$ bosons of $F_{+}\leq 0.10$ at the $95\%$ confidence level.}

\section{Introduction}

Since the discovery of the top quark in 1995 by the CDF and D\O~ collaborations~\cite{Abe,Aba}, the mass of this most massive known elementary particle has been measured with high precision. However, the measurements of other top-quark properties are still statistically limited, so the question remains whether the standard model successfully predicts these properties. 
In the following we present our measurement of the helicity fractions of $W$ bosons from top-quark decay.

At the Tevatron collider, with a center-of-mass energy of 1.96~TeV, most top quarks are pair-produced via the strong interaction. In the standard model the top quark decays in nearly 100$\%$ of all cases into a $W$ boson and a $b$ quark. Due to its large mass the top quark has a lifetime, that is shorter than the hadronization time. Thus, its decay products preserve the helicity content of the underlying weak interaction. The $V-A$ structure of the weak interaction in the standard model predicts that the $W^{+}$ bosons from the top-quark decay are dominantly either longitudinally polarized or left-handed, while right-handed $W$ bosons are heavily suppressed and even forbidden in the limit of a massless $b$ quark. 
Assuming a top-quark mass of 175~GeV$/c^{2}$ and neglecting the mass of the $b$ quark, the fraction of longitudinally polarized $W$ bosons is predicted~\cite{Kane} to be $F_{0}=0.7$, while the fraction of left-handed $W$ bosons is $F_{-}=0.3$. A significant deviation from the predicted value for $F_{0}$ or a nonzero value for the right-handed fraction $F_{+}$ could indicate new physics, such as a possible $V+A$ component in the weak interaction or other anomalous couplings at the $Wtb$ vertex.

The $W$ boson polarization manifests itself in the decay $W \rightarrow \ell\nu_{\ell}$ in the angle $\theta^{*}$, which is defined as the angle between the momentum of the charged lepton in the $W$ rest frame and the momentum of the $W$ boson in the top-quark rest-frame. The general $\cos\theta^{*}$ distribution is given by~\cite{Kane}:
\begin{equation}
\frac{dN}{d\cos\theta^*} \propto F_-\cdot\frac{3}{8}(1-\cos\theta^*)^2 + F_0\cdot\frac{3}{4}(1-\cos^2\theta^*) + 
F_+\cdot\frac{3}{8}(1+\cos\theta^*)^2
\label{eq:cosdistr}
\end{equation} \\

\section{Event Selection and Full Reconstruction}

In this analysis, we use the ``lepton+jets'' channel, where one $W$ boson originating from the top quarks decays leptonically into a charged lepton and a neutrino and the other $W$ boson hadronically into two quarks. Therefore, we select events with exactly one isolated electron or muon, substantial missing transverse energy due to the undetectable neutrino, and at least four jets. We require one of these jets to be tagged as $b$ jet, which means that it originates from a reconstructed secondary vertex, which is likely due to the long lifetime of $b$ hadrons. In the analyzed data set, we find 232 $t\bar{t}$ candidate events with a ``lepton+jets'' signature. 

In order to determine the $\cos\theta^{*}$ distribution for the selected events, the four-vectors of the top quarks, $W$ bosons, and of the charged lepton have to be reconstructed. The full reconstruction of the entire event starts with the reconstruction of the neutrino four-vector. The transverse components are obtained from the missing transverse energy, the $z$ component is calculated using a $W$ boson mass constraint. The $W$ boson four-vector is than obtained by adding the four-vector of the neutrino and of the charged lepton. Then all possibilities to assign the jets in the event to the two $b$ quarks from the top decay and to the two quarks from the hadronic $W$ decay are considered. This treatment leads to a multiplicity of hypotheses for the reconstruction of each event. For example, an event with four jets leads to 24 possibilities for the reconstruction.

The challenge is now to find the right hypothesis for each event. Therefore we make use of constraints on the mass of the reconstructed $W$ boson, on the mass difference between the two reconstructed top quarks, which should be zero within the resolution, and on the transverse energy of the two top quarks which should in leading order be equal to the transverse energy of the entire event. Finally, we prefer hypotheses, where jets tagged as $b$ jets are assigned to the $b$ quarks from the top decays. To estimate the probability for a tagged jet to be a real $b$ jet, we use a neural network $b$-tagger. Applying this method, for each single event we choose one hypothesis from which we then obtain the $\cos\theta^{*}$ distribution.
\begin{figure}
\begin{center}
\psfig{figure=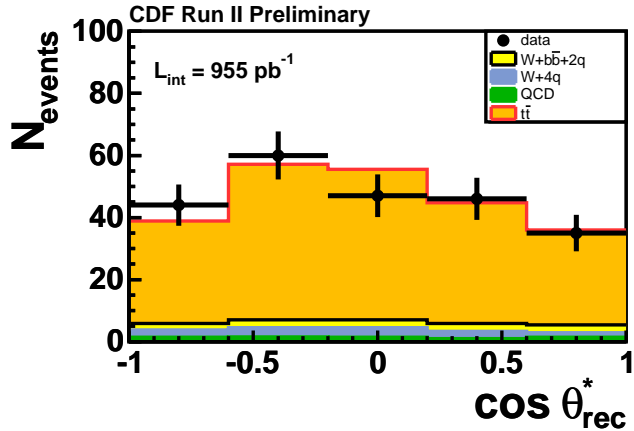,height=2.5in}
\caption{Measured $\cos\theta^{*}$ distribution together with the estimated signal and background.
\label{fig:dataraw}}
\end{center}
\end{figure}
Figure~\ref{fig:dataraw} shows the distribution of the measured $\cos\theta^{*}$ compared to the estimated signal and background distributions.

\section{Measurement}

Since the number of events in the data set is small, we do not simultaneously extract the fraction of longitudinally polarized and right-handed $W$ bosons. We either fix $F_{+}$ to zero and fit for $F_{0}$, or we fix $F_{0}$ to its expected value and fit for $F_{+}$. Thus, only one free parameter is used in each fit.

To extract the single free parameter ($F_{0}$ or $F_{+}$), we use a binned maximum likelihood method. The theoretically predicted number of events in each bin is the sum of the expected background and signal. The latter is calculated from the theoretical $\cos\theta^{*}$ distributions (see eq.~\ref{eq:cosdistr}) for the three helicities of the $W$ bosons.

Since the reconstruction of the $t\bar{t}$ process is not perfect, acceptance and migration effects have to be considered when calculating the number of signal events ($\mu_{k}^{sig,exp}$) expected to be observed in a certain bin $k$ after the reconstruction:

\begin{equation}
\mu_{k}^{sig,exp}\propto \sum_{i}\mu_{i}^{sig,theo}\cdot\epsilon_{i}\cdot S(i,k).
\label{eq:musigexp}
\end{equation} 

The migration matrix element $S(i,k)$ gives the probability for an event which was generated in bin $i$ to occur in bin $k$ of the reconstructed $\cos\theta^{*}$ distribution. Since the event acceptance depends on $\cos\theta^{*}$, we weight the contribution of each bin $i$ with the efficiency $\epsilon_{i}$. Both $\epsilon_{i}$ and $S(i,k)$ are determined using the standard model Monte Carlo generator PYTHIA~\cite{pyth}.

With the number of expected events and the number of observed data events in each bin, we minimize the negative logarithm of the likelihood function by varying the free parameter $F_{0}$ or $F_{+}$.

In addition, an upper limit for $F_{+}$ at the 95$\%$ confidence level is computed by integrating the likelihood function.

In order to compare our observations with theory, the background estimate is subtracted from the selected sample. To correct for the mentioned acceptance and reconstruction effects, we calculate a transfer function $\tau(F_{0},F_{+})$. The value $\tau_{i}$ for bin $i$ is given by the ratio of the normalized number of theoretically predicted events and the normalized number of events in this bin after applying all selection cuts and performing the reconstruction. For this calculation, we use the fit result of $F_{0}$ or $F_{+}$. Multiplying the background-subtracted number of events in bin $i$ with $\tau_{i}$ leads to the unfolded distribution, which then is normalized to the theoretically calculated $t\bar{t}$ production cross section~\cite{Ki,Ca} of $\sigma_{t\bar{t}}=6.7$~pb and can directly be compared with theory-curves for different values of $F_{0}$ and $F_{+}$ (see fig.~\ref{fig:Unfold}).

\section{Results}

\begin{figure}
\begin{center}
\begin{minipage}[c]{150mm}
\begin{center}
\subfigure[]{
\psfig{figure=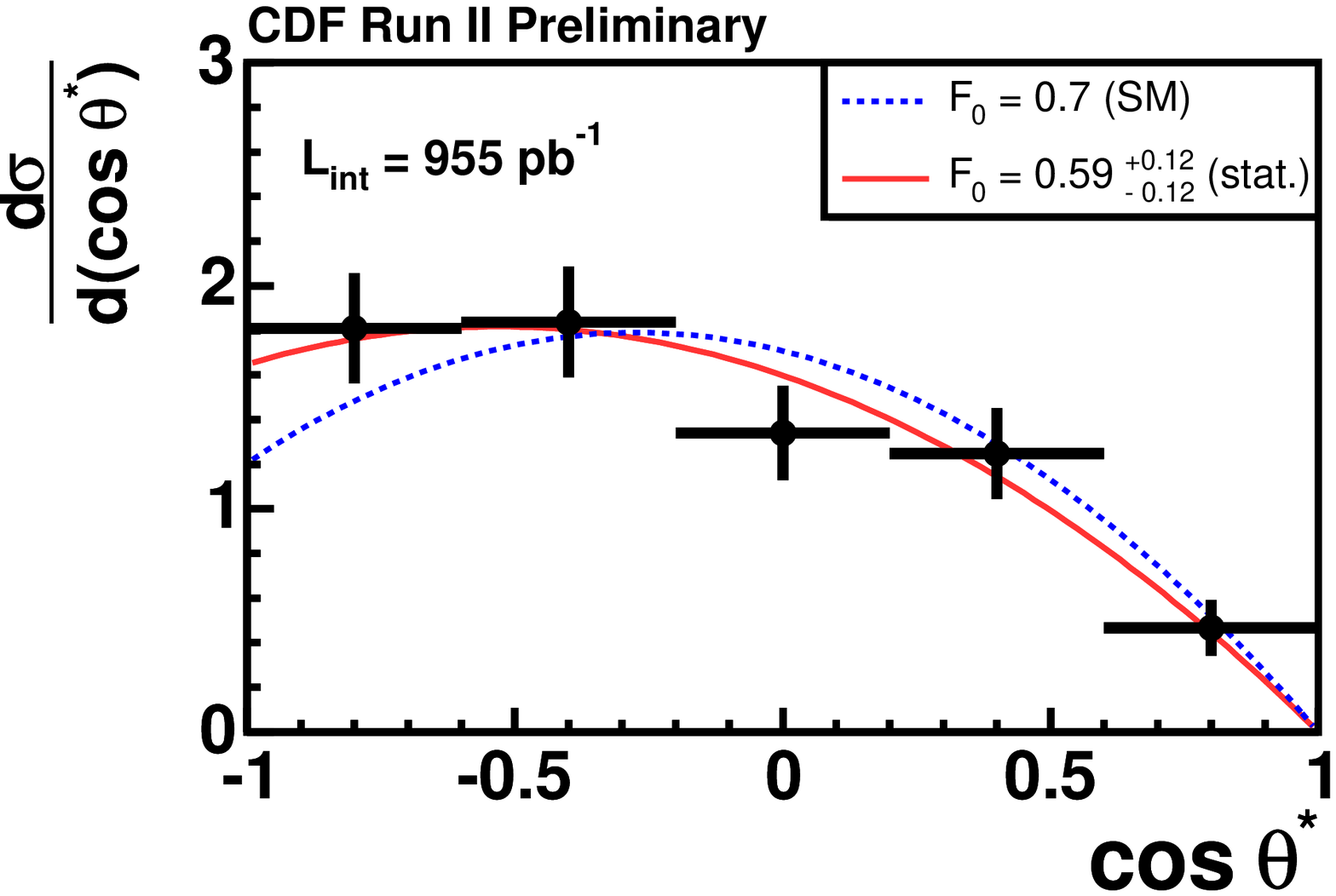,height=2.1in}}
\subfigure[]{
\psfig{figure=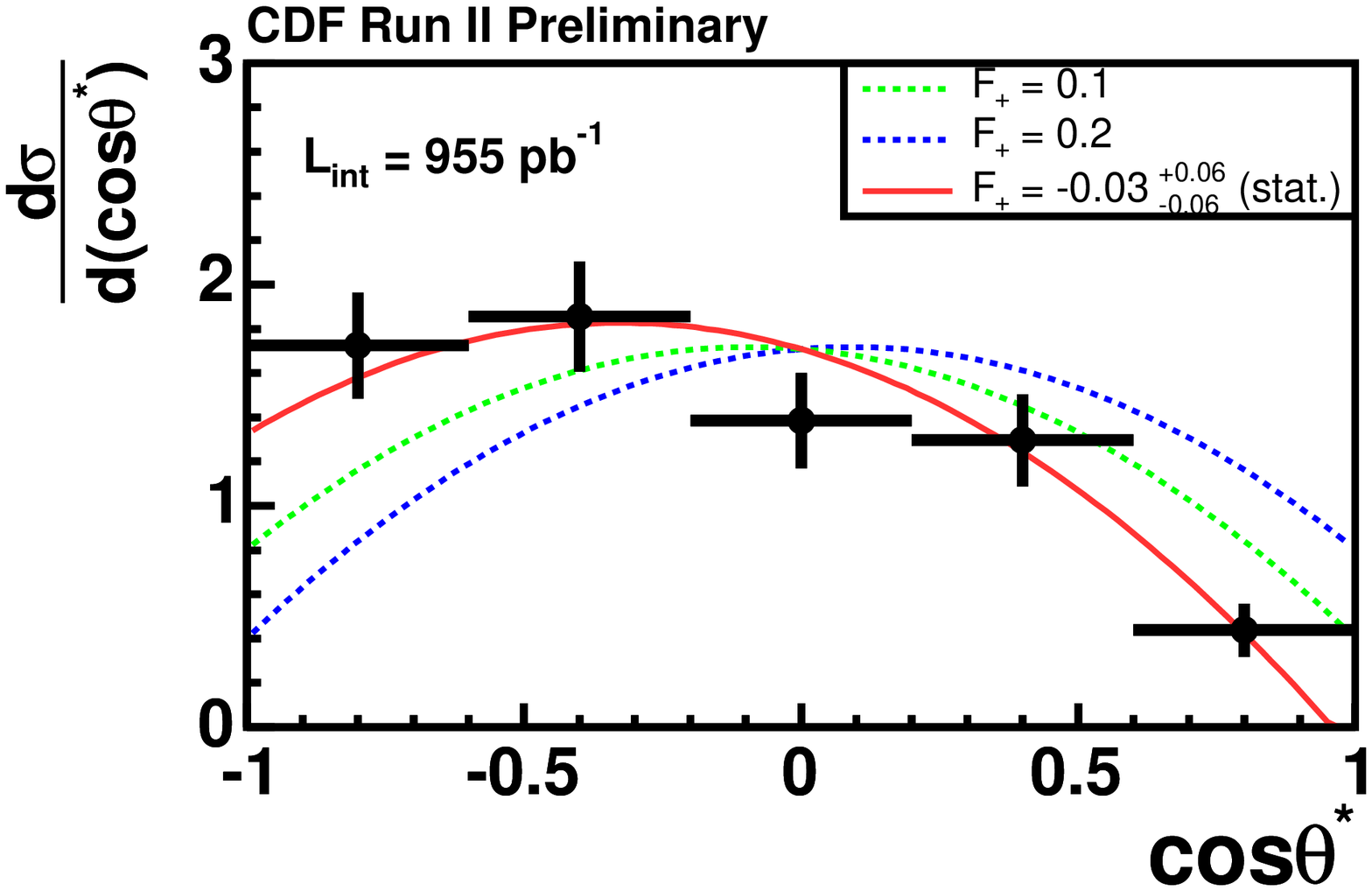,height=2.1in}}
\caption{Unfolded $\cos\theta^{*}$ distribution in data normalized to the theoretically calculated $t\bar{t}$ production cross section of $\sigma_{t\bar{t}}=6.7$~pb. (a) shows the unfolded distribution derived by applying the transfer function obtained from the fitted value of $F_{0}$. (b) displays the distribution unfolded with the transfer function obtained from the fitted value of $F_{+}$. In both plots the black points represent the corrected data, while the red smooth curves represent the theoretical distributions corresponding to the fitted values. The dashed lines represent distributions for different values of $F_{0}$ and $F_{+}$, respectively.
\label{fig:Unfold}}
\end{center}
\end{minipage}
\end{center}
\end{figure}

The data used in this analysis correspond to a total integrated luminosity of 955~$\rm{pb}^{-1}$, where 232 events have passed the event selection.
Taking systematic uncertainties into account, assuming a top-quark mass of 175~GeV$/c^{2}$, and assuming that the non-measured fraction is equal to its standard model expectation, the result for the fraction of longitudinally polarized and right-handed $W$ bosons is:
\begin{eqnarray}
  F_0 &=& 0.59\pm0.12~({\rm stat})~^{+0.07}_{-0.06}~({\rm syst})~, \nonumber\\
  F_+ &=& -0.03\pm0.06~({\rm stat})~^{+0.04}_{-0.03}~({\rm syst}).\nonumber 
\end{eqnarray} 

\noindent
We obtained an upper limit on the fraction of right handed $W$ bosons of $F_+ \leq 0.10$ at the 95\% confidence level.
Furthermore our method provides the possibility to correct the observed $\cos\theta^*$ distribution for the selected sample
for acceptance and resolution effects resulting in the distribution of the differential $t\bar{t}$ production cross section~(see figure~\ref{fig:Unfold}). As one can see, the observation is compatible with the curves predicted by the standard model. Also the measured values for $F_{0}$ and $F_{+}$ are within the uncertainties in good agreement with the standard model predictions.

\end{document}